\documentclass[a4paper,11pt]{article}

\usepackage{jinstpub} 
\usepackage{lineno}
\usepackage{url}
\usepackage{subcaption}

\begin{document}

\title{Generative Adversarial Networks for Scintillation Signal Simulation in EXO-200}

\author[1,a]{S.~Li}\note[a]{Corresponding author: sli135uiuc@126.com}\affiliation[1]{Physics Department, University of Illinois, Urbana-Champaign, Illinois 61801, USA}
\author[2,b]{I.~Ostrovskiy}\note[b]{Corresponding author: iostrovskiy@ua.edu}\affiliation[2]{Department of Physics and Astronomy, University of Alabama, Tuscaloosa, Alabama 35487, USA}
\author[3]{Z.~Li}\affiliation[3]{Physics Department, University of California, San Diego, La Jolla, CA 92093, USA}
\author[3]{L.~Yang}
\author[4]{S.~Al~Kharusi}\affiliation[4]{Physics Department, McGill University, Montreal, Quebec H3A 2T8, Canada}
\author[5]{G.~Anton}\affiliation[5]{Erlangen Centre for Astroparticle Physics (ECAP), Friedrich-Alexander-University Erlangen-N\"urnberg, Erlangen 91058, Germany}
\author[6,c]{I.~Badhrees}\note[c]{Permanent address: King Abdulaziz City for Science and Technology, Riyadh, Saudi Arabia}\affiliation[6]{Physics Department, Carleton University, Ottawa, Ontario K1S 5B6, Canada}
\author[7]{P.S.~Barbeau}\affiliation[7]{Department of Physics, Duke University, and Triangle Universities Nuclear Laboratory (TUNL), Durham, North Carolina 27708, USA}
\author[1]{D.~Beck}
\author[8]{V.~Belov}\affiliation[8]{Institute for Theoretical and Experimental Physics named by A.I. Alikhanov of National Research Centre ``Kurchatov Institute,'' Moscow 117218, Russia\footnote{Now a division of National Research Center ``Kurchatov Institute,'' Moscow 123182, Russia}}
% ...

\abstract{Generative Adversarial Networks trained on samples of simulated or actual events have been proposed as a way of generating large simulated datasets at a reduced computational cost. In this work, a novel approach to perform the simulation of photodetector signals from the time projection chamber of the EXO-200 experiment is demonstrated. The method is based on a Wasserstein Generative Adversarial Network — a deep learning technique allowing for implicit non-parametric estimation of the population distribution for a given set of objects. Our network is trained on real calibration data using raw scintillation waveforms as input. We find that it is able to produce high-quality simulated waveforms an order of magnitude faster than the traditional simulation approach and, importantly, generalize from the training sample and discern salient high-level features of the data. In particular, the network correctly deduces position dependency of scintillation light response in the detector and correctly recognizes dead photodetector channels. The network output is then integrated into the EXO-200 analysis framework to show that the standard EXO-200 reconstruction routine processes the simulated waveforms to produce energy distributions comparable to that of real waveforms. Finally, the remaining discrepancies and potential ways to improve the approach further are highlighted.}


\maketitle

\flushbottom

\section{Introduction}
\label{sec:intro}
Computer simulations play a crucial role in many aspects of experimental nuclear and particle physics, including detector design optimization, data analysis, and new physics searches. A typical simulation uses a Monte Carlo approach that starts with a generation of a primary particle that is then propagated through the detailed detector geometry taking into account the stochastic nature  of relevant physics processes and detector responses. For simulation of the response to scintillation light produced by ionizing radiation, tens of thousands of photons need to be generated for each O(MeV) energy deposition in liquid xenon. Since the trajectory of each photon is tracked throughout its propagation by simulation packages, such as Geant4\cite{g4}, the process is computationally expensive and time consuming. Recent development has significantly sped up the scintillation light simulation through the use of software packages that utilize graphical processing units (GPUs), like Chroma\cite{chroma_wp,sorting,chroma_darwin,nexo}, but even these highly parallel simulations still require large computational resources. Another issue is that optical properties of materials and detector geometry are often not precisely known, which contributes to differences between the simulation and experimental detector response.  

Recent developments in machine learning based generative models offer alternative approaches to physics event generation, including Generative Adversarial Networks (GANs)~\cite{martin2017wasserstein}, Variational Autoencoders (VAEs)~\cite{KingmaAuto-Encoding}, and Normalizing Flows (NFs)~\cite{9089305}. These techniques allow for non-parametric learning of the data distribution, with sampling being as fast as a single forward pass through the neural network. GAN was invented for computer image generation. It consists of two competing networks, a generator and a discriminator. While the generator does its best to mimic the training images, the discriminator aims to separate the real images from the generated image~\cite{goodfellow2014generative}. Successful training of the GAN network can lead to the generation of new images indistinguishable from the training images. An obvious caveat is that training images must be available, either from the specific detector to be studied or from essentially similar detectors. Several groups have demonstrated GANs as a tool for fast simulation of Cherenkov detectors~\cite{maevskiy2020fast}, muon production through the interaction of a proton beam with dense targets~\cite{ahdida2019fast}, liquid argon time projection chambers~\cite{tuftsDL}, and high-granularity calorimeters \cite{calorimeter-2022}.

In this work, we apply the GAN technique to the simulation of scintillation light in the EXO-200 experiment~\cite{auger2012exo}. EXO-200 is a 175-kg liquid xenon (LXe) detector built to search for the neutrinoless double beta decay of $^{136}$Xe. The scintillation light is collected by two planes of avalanche photo-diodes (APDs). While some progress was made on developing a detailed optical simulation of the EXO-200 detector, the discrepancies between data and simulation, likely caused by poorly known optical properties, and computational costs of photon tracking through a complex detector geometry led to EXO-200 using a simplified, parametric optical simulation of the overall light yield per one array of APDs. In this work, we demonstrate that one can train a GAN network directly with calibration data from EXO-200, bypassing the needs for detailed knowledge of optical properties and detector geometry. Once well-trained, the generator is able to produce accurate response of individual APDs at given positions and energies, with better fidelity and faster speed compared with conventional MC simulation. Section~\ref{sec:tpc} provides a brief overview of the EXO-200 detector, its simulation and event reconstruction. Section~\ref{sec:models} describes the GAN models developed for generating raw signals deposited on APD channels, including section~\ref{sec:training} that explains our approach to simulating APD signals with GANs.
Results of the simulations are presented in Section~\ref{sec:results}. The last section summarizes the findings and discusses the limitations and future directions of the work.

%%%%%%%%%%%%%%%%%%%%%%%%%%%%%%%%%%%%%%%%%%%%%%%%%%%%%%%%

\section{The EXO-200 experiment}
\label{sec:tpc}
\begin{figure}
     \centering
     \includegraphics[width=0.6\textwidth]{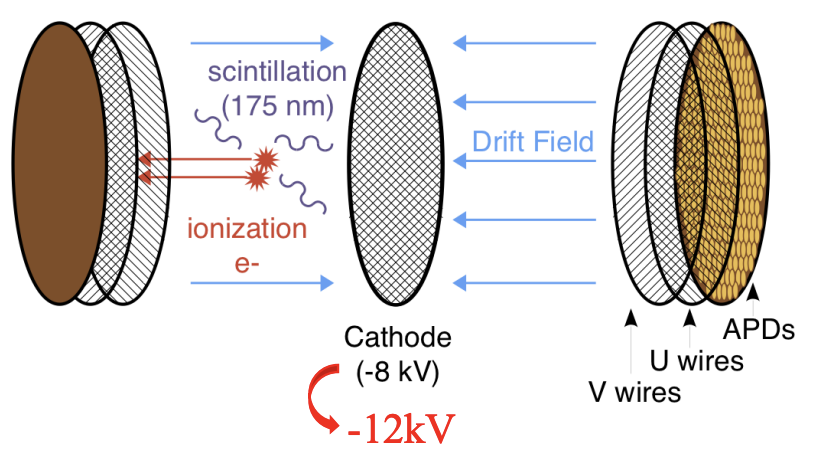}
     \caption{Schematic of the inner EXO-200 TPC. The cathode voltage was set to -8kV (-12kV) for Phase I(II) operations. The ionization electrons are drifted to the U wire plane that is grounded and serves as an anode. The scintillation light of $\sim$175 nm~\cite{lxe_wave} is detected by APDs behind the grids.}
     \label{a269_fig:TPC}
\end{figure}

The EXO-200 detector is a cylindrical time projection chamber containing 175 kg of LXe enriched to 80.6\% in $^{136}$Xe. The chamber is divided into two equal drift volumes by a photo-etched phosphor bronze cathode plane in the middle, as shown in Fig.~\ref{a269_fig:TPC}. At each end of the chamber there are two instrumented wire planes crossed at 60$^\circ$ to measure charge. When ionizing radiation interacts in liquid xenon, the produced electrons are drifted towards the wire planes by a main drift field applied between the wires and the cathode plane. The V wires are located closer to the cathode and measure the induction of the passing electrons, which are collected onto the U wires that are located 9~mm behind. The U- and V-wire planes together have 91.8\% optical transparency. The APDs are located on a platter behind the wire planes looking at the drift volume. The APDs combine high quantum efficiency for the scintillation light with ultra-low levels of radioactivity~\cite{neilson2009characterization}. The wire signals provide two-dimensional position information of the event. The position in the drift direction can be obtained using the known drift speed and the time difference between the light signal collected by the APDs and charge signals collected by the U wires. Given the drift speed and the size of the detector, the maximum drift time is on the order of 120 \textmu s. The charge and light signals in LXe are anti-correlated~\cite{anticor,anticor2}. Consequently, the two signals are combined for energy measurements to achieve optimal energy resolution. Both the wire and APD signals are read out by charge-sensitive preamplifiers outside the lead shielding. A detailed description of the detector can be found in~\cite{auger2012exo}.

When a detector trigger condition is met, the data acquisition (DAQ) system records digitized raw waveforms of 152 charge channels and 74 light channels. Each waveform consists of 2048 samples taken at 1 MS/s with 1024 samples occurring before the trigger and 1024 samples after the trigger. The raw waveforms are processed by an algorithm that reconstructs the energy depositions inside the detector. Briefly, an initial signal-finding stage identifies channels containing a signal above a given noise threshold. After the signal-finding stage is the parameter estimation stage, where each of the identified signals is analyzed to extract parameters relevant for the analysis. Then, the set of signals and signal parameters for each event is combined to determine event topology and event energy. The reconstructed signals are calibrated by radioactive sources emitting $\gamma$ rays with known energy. The charge and light signals are calibrated separately, then combined to form a rotated energy scale with improved energy resolution. 

To understand the detector response to energy deposits in the detector volume, a Geant4-based Monte Carlo simulation is employed. The simulation can accurately simulate the charge depositions. It does not produce an accurate response for individual APD channels because of the uncertainty of optical properties and difficulties with implementing the anti-correlation between charge and light responses. Instead, it avoids the computational cost of tracking individual photons by using a parameterized light response function to simulate the light yield on each plane of the APDs based on the position of the energy deposit in the detector. The light response is then evenly distributed among all APD channels of the given plane with randomized noise added to each waveform. This light simulation is used only to approximately simulate the light reconstruction threshold and is not used to determine the Monte Carlo energy. Following other successful applications of Chroma to optical simulations~\cite{chroma_wp,sorting}, a Chroma-based photon tracking was later developed in EXO-200. This approach matched with the data better but still required tuning of material optical properties and has not yet been published. The difficulties with the scintillation light simulation motivate us to apply the GAN approach to improve the simulation quality and speed. 
%%%%%%%%%%%%%%%%%%%%%%%%%%%%%%%%%%%%%%%%%%%%%%%%%%%%%%%%
\section{GAN Description}
\label{sec:models}

\subsection{APD waveform image}
Since the GAN framework was developed for image generation, we convert the 74 channels of APD digitized waveforms into images for ease of integration into the software framework. We remove the baseline of the waveforms by subtracting the average of the first 300 \textmu s. We do not scale the waveforms, as it did not improve training. A typical waveform's values range between -100 to 500 ADC units. To reduce the number of dimensions of the target space, we select 350 \textmu s samples of the waveform around the signal. Since the rest of the samples only contain noise, this achieves a substantial reduction in complexity without an appreciable reduction in accuracy. The location of the signal depends on the trigger type -- at the center of the waveform for events triggered by APDs and off-center but within the maximal drift time for events triggered by wires.
\begin{figure}[hpb]
    \centering
    \includegraphics[width=\textwidth]{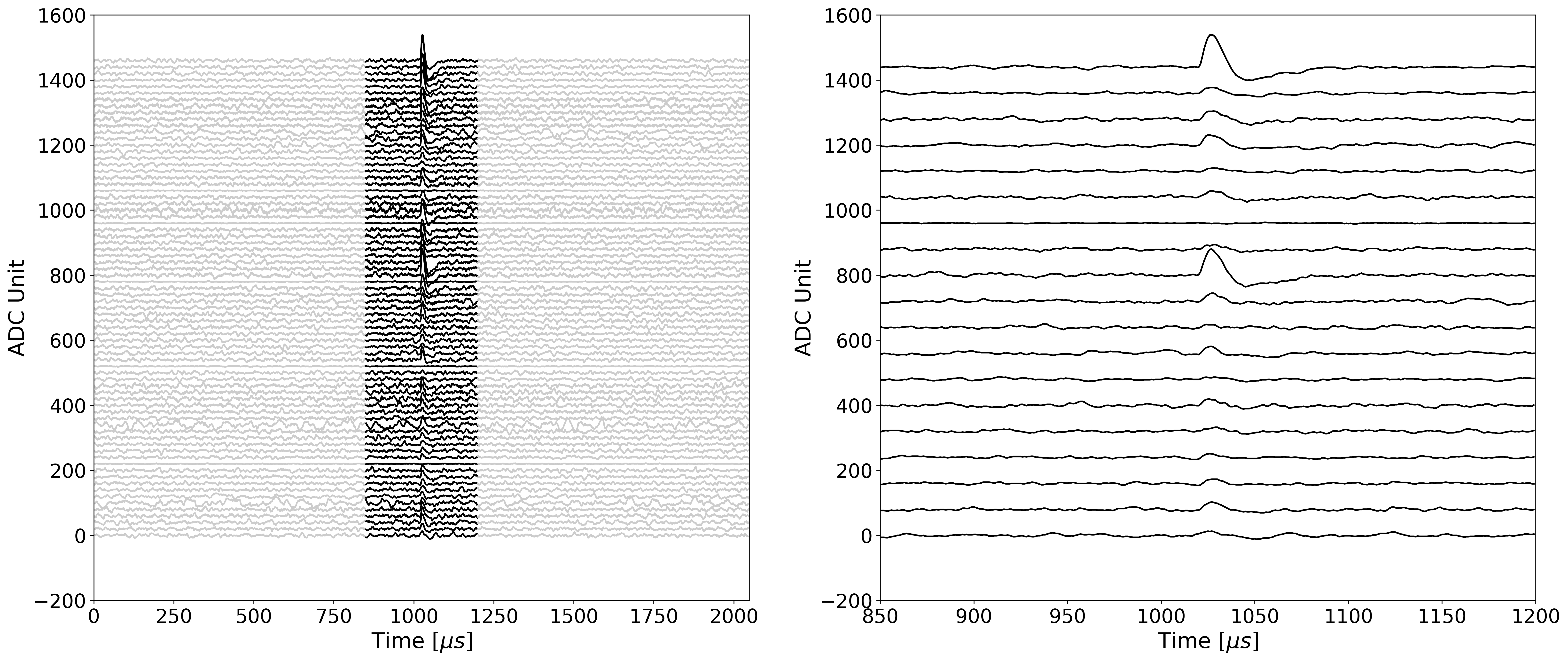}
    \caption{(Left) Example waveforms of the APDs showing pulses from a cluster at certain position and energy. For clarity, the signals on each APD are offset vertically in order of channel number and separated by a constant of 20 units in the y direction. Due to computational reasons, raw waveforms of 2048 \textmu s (gray) are truncated to 350 \textmu s (black) before being fed to the neural network. (Right) Same as the left panel but zoomed-in around the peak positions and showing every fourth APD channel for clarity.}
    \label{fig:wf-samples}  
\end{figure}
The resulting image of an event consists of 74 channels, each having 350 entries, as shown in Fig.~\ref{fig:wf-samples}. 

\subsection{Network structure}
To simulate the raw waveforms of APDs, we use the Keras framework~\cite{chollet2015keras} to build a network based on Wasserstein GAN~\cite{martin2017wasserstein} with label conditioning. The network consists of a discriminator and a generator. The generator takes in an array of uniform random values, specified position and energy of the scintillation cluster as inputs, then generates artificial waveforms mimicking the real APD responses. The discriminator measures the similarity between the generated and real samples, providing feedback to the generator to improve its performance. Training of such an adversarial framework allows the generator to produce data-like samples $x_g =G(z)$ out of noise $z$. The original GAN~\cite{goodfellow2014generative} architecture has shown impressive results but turned out to be unstable and hard to monitor during the training process. The Wassertein GAN is, with its improvement, a network that allows for a stabilized training procedure by delivering adequate gradients to the generator, which provide a meaningful loss metric not susceptible to training collapse. The Wassertein GAN is enforced by the Lipschitz constraint on the discriminator to calculate the 1-Wasserstein (or, simply, Wasserstein) distance. A differentiable function $f$ is 1-Lipschitz if and only if it has gradients with norm at most 1 everywhere, $\forall x:\; |f'(x)| \leq 1$. A later study~\cite{gulrajani2017improved} proves that points interpolated between the real and generated data should have a gradient norm of 1 for $f$. Hence, instead of applying weight clipping which manually forces hyperparameters of network in the Lipschitz constraint, Wassertein GAN gradient penalty (WGAN-GP) penalizes the model if the gradient norm moves away from its target norm value of 1, as shown in Eq.~\ref{equ:GP} from~\cite{gulrajani2017improved}. 
\begin{equation}
L = \underbrace{\underset{\tilde{x}\sim P_g}{\mathbb{E}}[D(\tilde{x})] - \underset{x\sim P_r}{\mathbb{E}}[D(x)]}_{\text{Wasserstein\,distance}}+ \underbrace{\lambda \underset{\hat{x} \sim P_{\hat{x}}}{\mathbb{E}} [(\lVert \nabla D(\hat{x}) \rVert_2 - 1)^2]}_{\text{gradient\,penalty}}
\label{equ:GP}
\end{equation}
where $D$ is the discriminator, $ \hat{x} = \epsilon x + (1 - \epsilon ) \tilde{x}, \epsilon \in U(0,1)$, $\lambda$ is the gradient penalty's weighting coefficient, and $\lVert \rVert_2$ denotes the Euclidean norm. The gradient penalty term, $(\lVert \nabla D(\hat{x}) \rVert_2 - 1)^2$, encourages the norm of the gradient to go towards 1. The point $x$ used to calculate the gradient norm is any point sampled between the GAN-generated distribution, $P_g$, and real data distribution, $P_r$. A gradient penalty is a soft version of the Lipschitz constraint that removes the undesirable behaviour of gradient explosion/vanishing when the weight clipping parameter is not carefully tuned in the earlier Wassertein GAN design. 

%%%%%%%%%%%%%%%%%%%%%%%%%%%%%%%%%%%%%%%%%%%%%%%%%%%%%%%%%%%%%%%%%%
Wasserstein Distance is also known as Earth mover’s distance, as it defines the cost for moving a distribution onto a target distribution using optimal transport. The original GAN is trained to minimize the Jensen-Shannon Divergence (JD)~\cite{goodfellow2014generative}. Comparing with JD, the Wasserstein Distance has the following advantages:
\begin{itemize}
    \item Wasserstein Distance is a continuous and almost differentiable function which is easier to optimize.
    \item As the discriminator gets better, JD locally saturates and thus the gradient becomes zero and vanishes.
    \item Wasserstein Distance is a meaningful function as its converges to 0 while two distributions are getting closer together and diverges when they are moving apart.
    \item Wasserstein Distance is more stable than JD, and the model is hard to collapse when using Wasserstein Distance as the objective function.
\end{itemize}   %\textcolor{red}{do we need references on this? }
%%%%%%%%%%%%%%%%%%%%%%%%%%%%%%%%%%%%%%%%%%%%%%%%%%%%%%%%%%%%%%%%%%
Lastly, to generate samples with specific characteristics, in this case, the position and energy of the scintillation, the generator is provided with the labels through label conditioning by an Auxilliary Classifier, as described in~\cite{odena2017conditional}. Using this Generator model, the training process becomes more stable and can now be used to generate images of a specific type using the class label.
 
\begin{figure}
    \centering
    \includegraphics[width=0.6\textwidth]{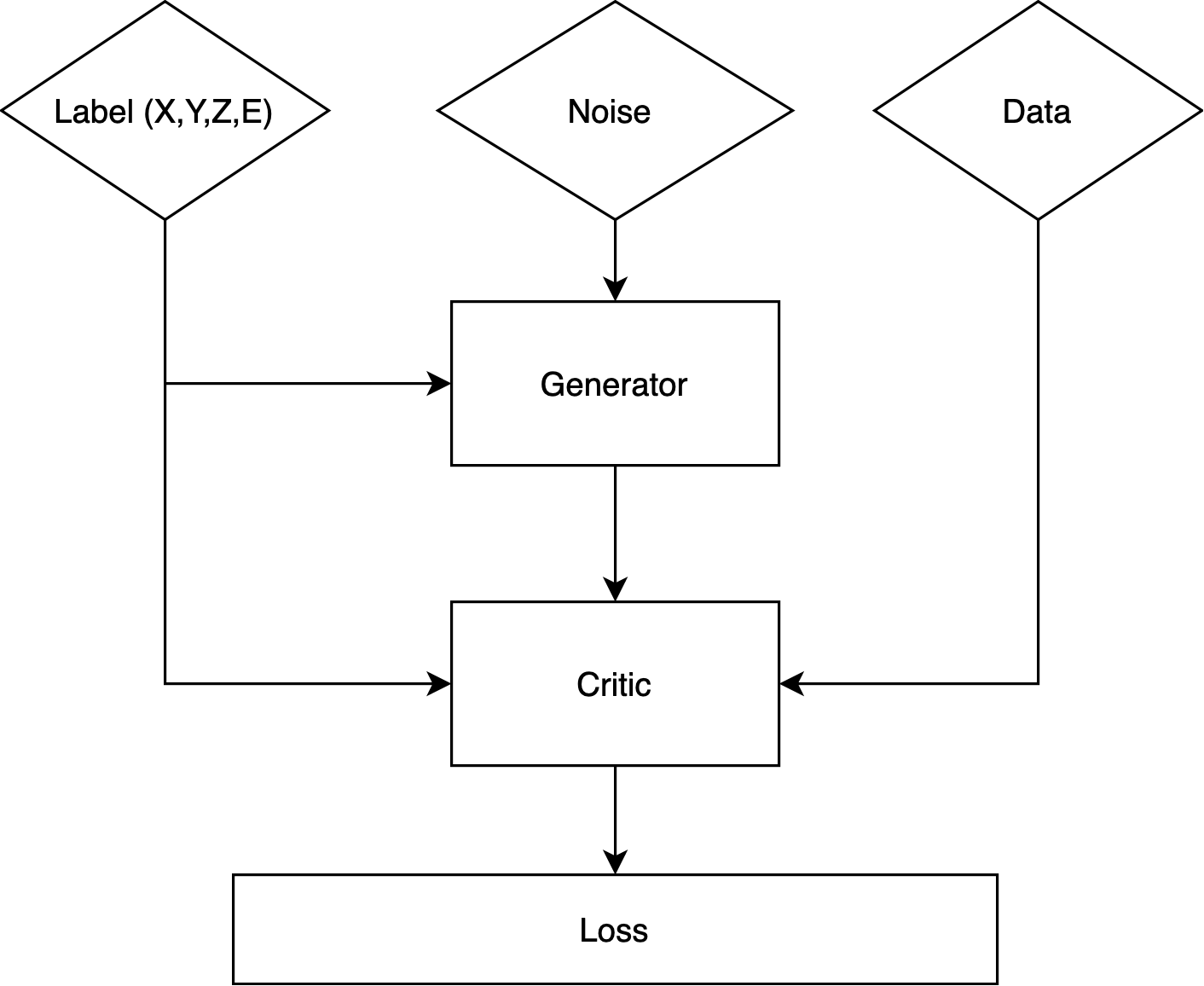}
    \caption{The schematic training diagram of generator and discriminator for the GAN network. Labels, (X,Y,Z,E), are also inputs of discriminator to allow training of generator with specific characteristics.}
    \label{fig:architecture}
\end{figure}

The schematic diagram in Fig. \ref{fig:architecture} shows the overall structure of the GAN model. The detailed layer structure of the generator and the discriminator network can be found in Appendix~\ref{appendix_A}.
The generator dependency can be written as $G = G(N, E, P)$, where $N$ is random noise and $E$ and $P$ are physics labels corresponding to scintillation energy E and the deposit coordinates $P = (P_x,P_y,P_z)$. Because we use calibration data for training the network, we do not know the precise energy and position of the scintillation cluster, and need to use the EXO-200 reconstruction algorithm to provide labels for the events. There are two choices for the scintillation energy label, one is the reconstructed scintillation energy and the other is a combined, also called rotated, reconstructed energy that uses both charge and light information. Although the rotated energy has better energy resolution of $\sim$1.2\% at the Q value of the $^{136}$Xe double-beta decay~\cite{exo_final}, it removes the anti-correlation between the charge and light responses and so is a biased estimate of an event's true scintillation energy. The reconstructed scintillation energy, despite its larger energy resolution of $\sim$5.1\% at the same energy, is thus a better estimate of the true scintillation energy and is chosen as the energy label for the data in this work. 

\begin{figure}
    \centering
    \includegraphics[width=0.5\linewidth]{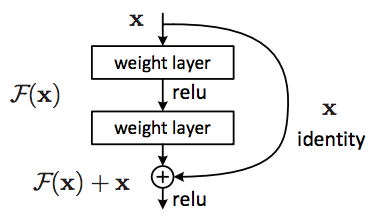}
    \caption{A block of residual learning \cite{he2016deep}. }
    \label{fig:residual}
\end{figure}

The physics labels are tiled to the convolutional layers as shown in Fig~\ref{fig:GAN_networks} in Appendix~\ref{appendix_A}.
Each value of the input dimensions, $x_i$, is copied to a new convolutional layer filled with that value, so four additional layers are concatenated to the image input, changing the dimension from (74,350,1) to (74,350,5).   
Furthermore, we also concatenate the label vector to the dense representation layers at the end of the convolutional part of the discriminator network. This design is inspired by the concept of residual learning~\cite{he2016deep} which simplifies learning when the output is expected to be close to input.
The basic structure of residual learning is shown in Fig.~\ref{fig:residual} in which the weight layer only needs to learn the difference between input and output.

%%%%%%%%%%%%%%%%%%%%%%%%%%%%%%%%%%%%%%%%%%%%%%%%%%%%%%%%
\subsection{Training and test datasets}
\begin{figure}
    \centering
    \includegraphics[width=0.8\linewidth]{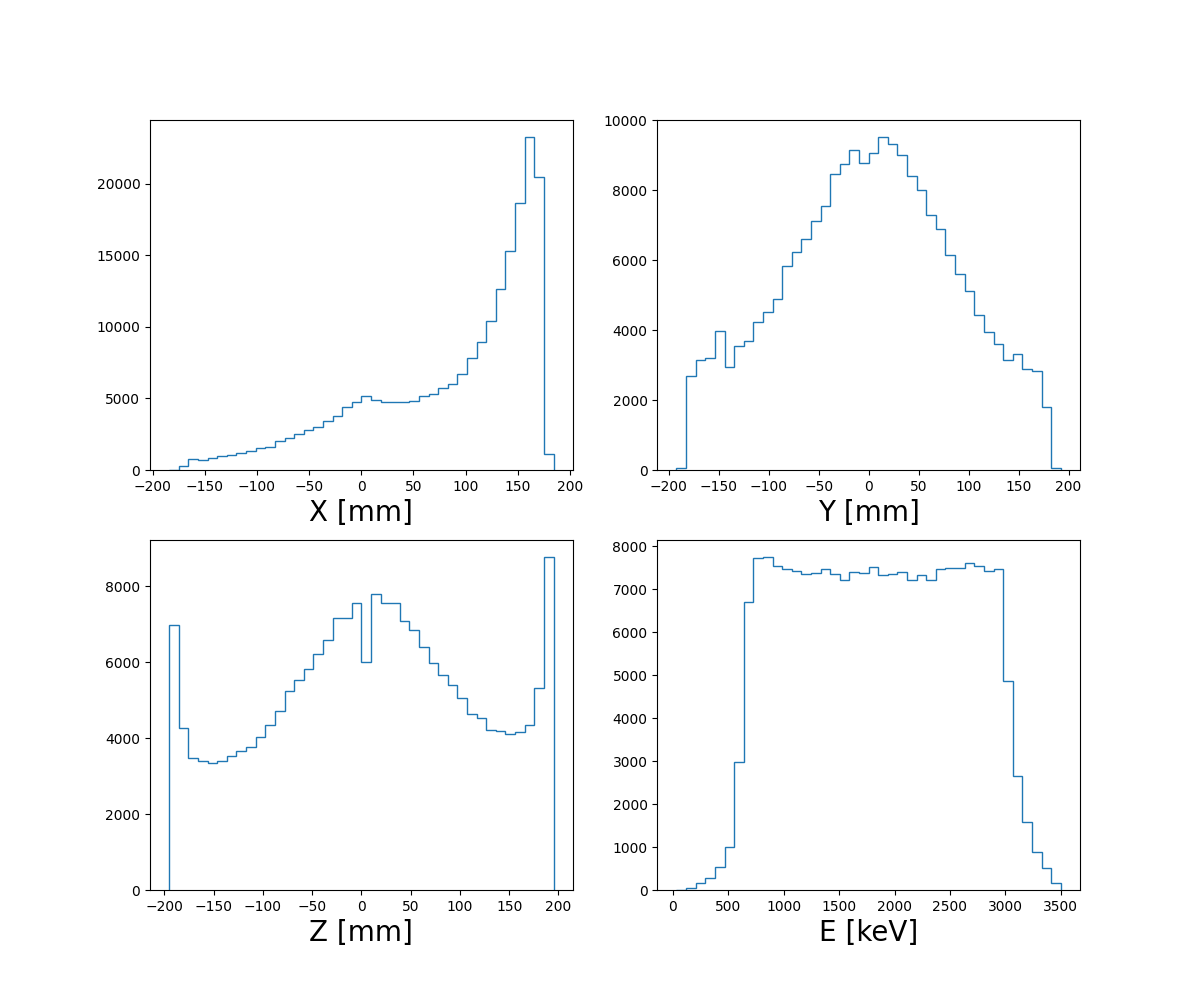}
    \caption{The spatial and energy distribution of training dataset.}
    \label{fig:train_set}
\end{figure}

The training dataset is obtained from $^{228}$Th calibration runs of the EXO-200 experiment.  
The data are separated into two phases: phase I, which took place from 2011 to 2014, and phase II (2016 to 2018). The noise of the APD channels is different in the two phases because the frontend electronics were upgraded at the start of phase II. In this work, we focus on the simulation of the phase II data. The training set is created by randomly picking events from the calibration runs. The events are selected to only have one scintillation and one charge cluster, with fully reconstructed energy and position. The training objects are created by combining the waveforms of all individual APD channels to form an image. The pixel values of the image correspond to the waveforms’ amplitudes at a particular time. The reconstructed energy and position serve as event labels. To reduce bias, the energy distribution of the training sets is flattened. The same method was applied to flatten the spatial distribution of events. However, since the calibration source positions are limited, and certain regions of the detector have very few events due to a limited amount of nearby source deployments, we cannot completely flatten the training dataset in the spatial dimension, as shown in Fig.~\ref{fig:train_set}. 
The final training set contains 215,789 events from phase II. Similarly, a test dataset is obtained in the same way after event selection for the training dataset. The ratio of the total number of events of test and training datasets is close to 7:3.

%%%%%%%%%%%%%%%%%%%%%%%%%%%%%%%%%%%%%%%%%%%%%%
\subsection{Network Training Process}
\label{sec:training}
Because a GAN contains two separate networks, its training algorithm must deal with two different kinds of training. Typically, the algorithm proceeds in alternating training of the discriminator and the generator. During training of the discriminator, the real data and the generated data serve as inputs. Then the training switches to the generator. The generator performance improves with the discriminator providing the feedback. The cycle continues until the Wasserstein distance becomes close to zero, indicating that the generator produces objects that the discriminator cannot effectively distinguish from the real ones. The convergence of the GAN network can be difficult to achieve and identify, as it requires a delicate balance between the discriminator and the generator networks. 

In this work, we train both generator and discriminator using the Adam optimizer~\cite{kingma2014adam} with the learning rate starting at 10$^{-4}$ at the beginning of the training process and with the optimizer parameters $\beta_1$ = 0.5 and $\beta_2$ = 0.9. The loss function coefficient number $\lambda=10$. We train the discriminator for five iterations, before updating the generator for one iteration. In each iteration step, the network is trained with a 20 events batch. One epoch counts as a loop over the entire phase II training set. We train the networks on a computing cluster in successive batch jobs running for 10 epochs each, while decreasing training rate gradually from 10$^{-4}$ to 10$^{-8}$ for the discriminator and 5 times faster for the generator.

\begin{figure}
    \centering
    \includegraphics[width=0.75\textwidth]{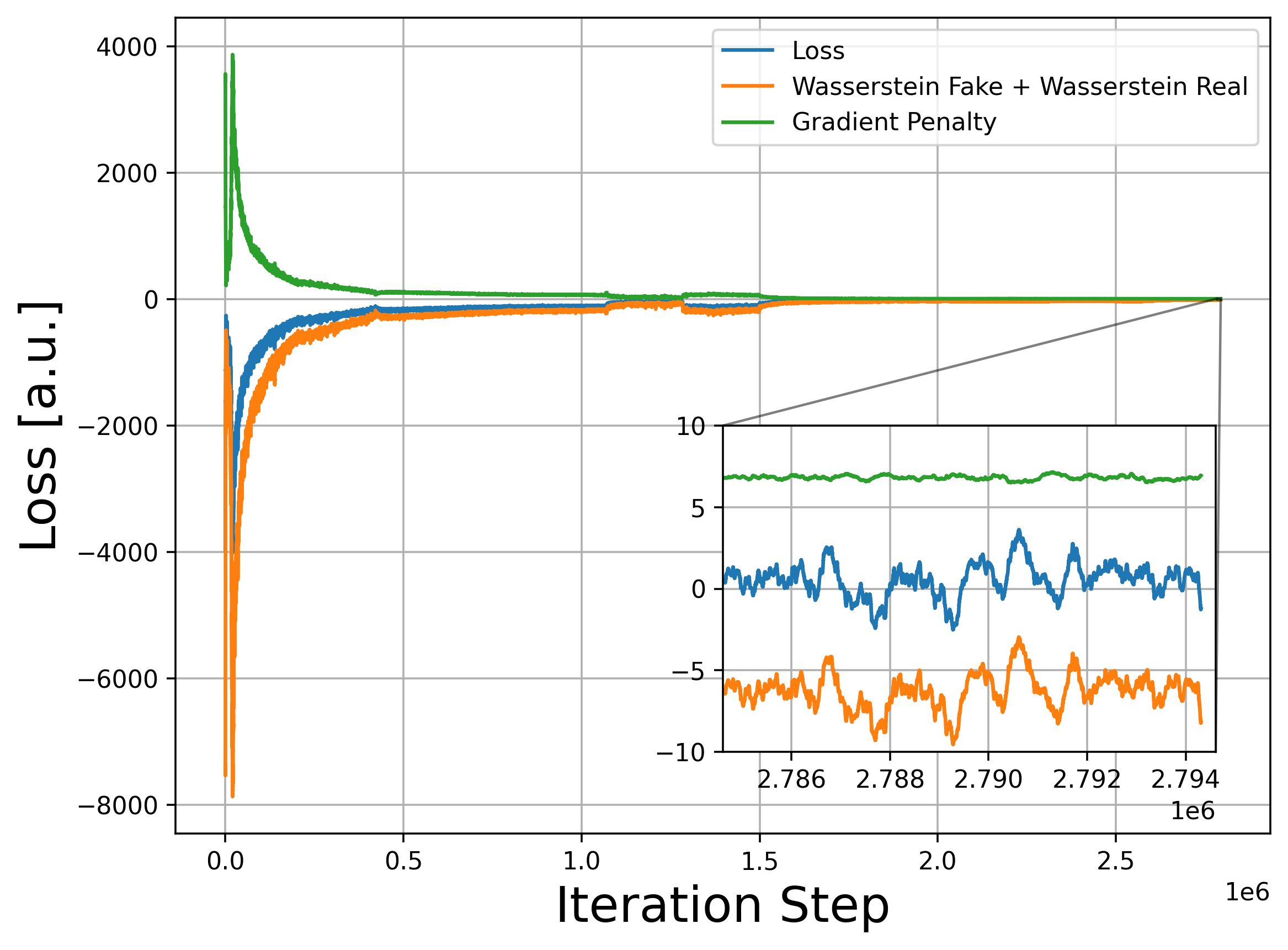}
    \caption{Loss curves of the moving average of 300 iterations for total discriminator loss during the training (blue), the Wasserstein distance (orange) and gradient penalty (green). The Wasserstein term contains "wasserstein real" as $-D(x)$, the outputs of real waveforms and "wasserstein fake" as $D(G(z))$, the outputs from GAN-generated waveforms.}
    \label{fig:critic_history}
\end{figure}

The gradient penalty and Wasserstein distance drop quickly during the initial training, but the rate of decrease slows down and gradually approaches zero, as shown in Fig.~\ref{fig:critic_history}. The training results were evaluated during the training using the metrics described in the next section. We stop the training if absolute values of the gradient penalty and Wasserstein distance are less than 10, and the loss is close to zero, which typically occurs after 2.8 million batches, or about 260 epochs. The training process takes about 500 GPU-hours on the Nvidia GeForce 1080Ti GPU with 11 GB RAM.

%%%%%%%%%%%%%%%%%%%%%%%%%%%%%%%%%%%%%%%%%%%%%%%%%%%%%%%%%%%%%%%%%%
\section{Results and validation}
\label{sec:results}
%%%%%%%%%%%%%%%%%%%%%%%% Explanation of evaluation %%%%%%%%%%%%%%%%
For evaluating the model, we pair up the events from the test dataset with the ones generated with the GAN for the same energy and position parameters. The generated events not only need to ``look similar" but also need to have similar features such as the energy and position information after reconstruction. Thus we introduce a set of metrics including the raw waveforms, the features extracted from the pulses, such as amplitudes and spatial dependency, and reconstructed physical information, such as scintillation and rotated energy.
%%%%%%%%%%%%%%%%%%%%%%%%%%%%%%%%%%%%%%%%%%%%%%%%%%%%%%%%%%%%%%%%%%%%
The validation tests for the training results are conducted in following levels:
\begin{itemize}
    \item Raw waveform level: directly compare raw waveforms 
    \item Signal level: compare signal features and its spatial dependence
    \item Reconstruction level: compare information from the EXO-200 reconstruction, e.g., scintillation energy spectrum.
\end{itemize} 

\begin{figure}[htpb]
        \centering
        \includegraphics[width=\linewidth]{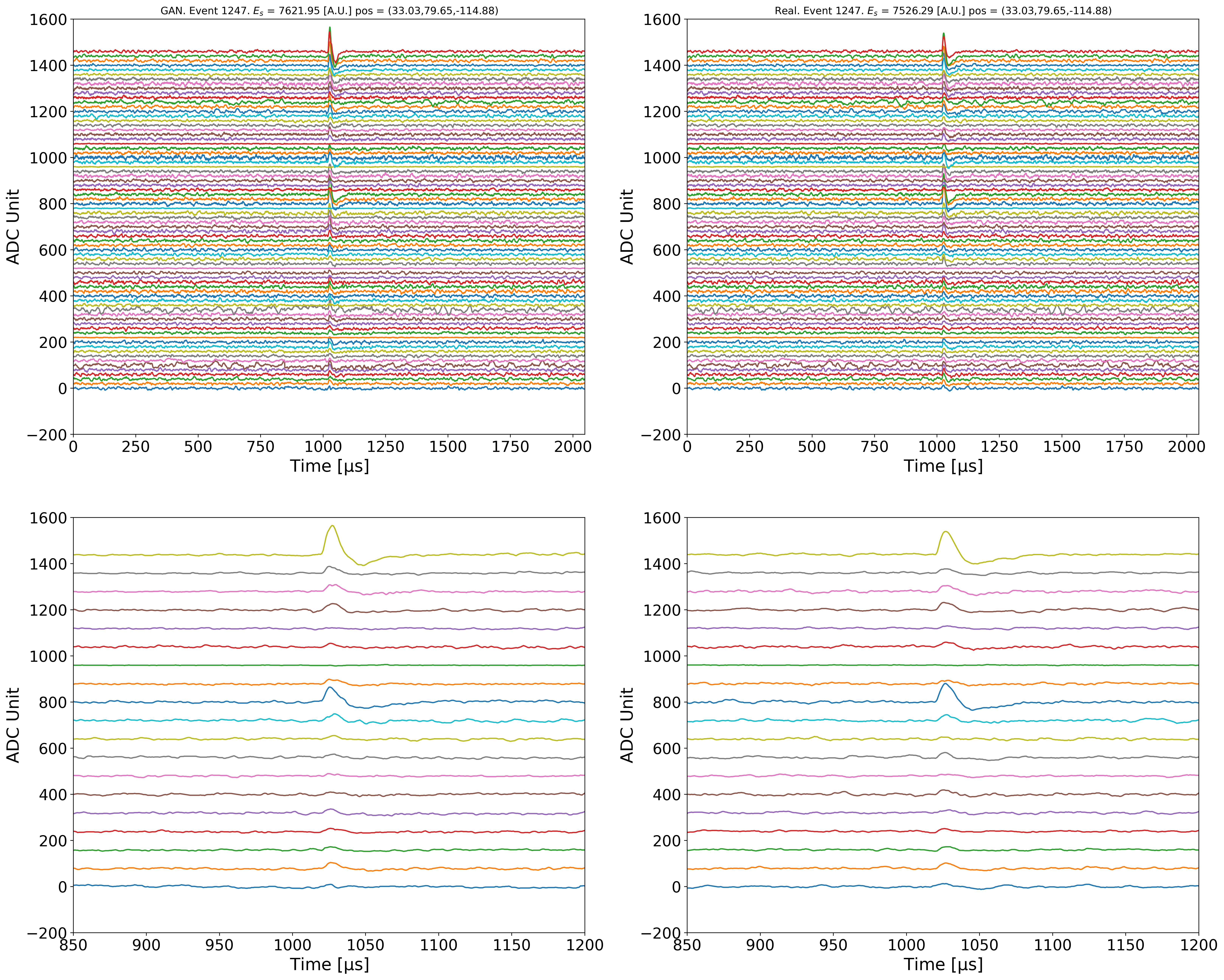}
        \label{fig:wf_sample_B_high}
    \caption{Example of the generated (top left) and real (top right) APD signal waveforms. Vertical and horizontal axes correspond to the amplitude + channel number $\times$ A.U. and time bins, respectively. The waveforms are obtained for the same values of the conditional variables and look qualitatively very similar. The two bottom panels are the same as the top ones but zoomed-in around the peak positions and showing every fourth APD channel for clarity.}
    \label{fig:wf_sample_B}
\end{figure}

Because the length of the GAN-generated waveforms is 350 \textmu s, shorter than the full EXO-200 event size of 2048 \textmu s, they are stitched back to the full waveform with the section centered around the event trigger replaced by the generated waveforms. We visually compare the generated and real events in different energy ranges, low ($<$1 MeV), middle (1 -- 2 MeV), and high ($>$2 MeV), and find that the generated waveforms reproduce the main features of the real events. The signals in each of the waveforms are aligned in time as one would expect from scintillation events. The signal shapes, channel noise, and amplitude distribution in the generated image is also similar to the real events. Fig.~\ref{fig:wf_sample_B} shows a sample event with energy $>$2 MeV. 
 
\begin{figure}[hptb]
\centering
    \begin{subfigure}{.95\textwidth}
    \centering
    \includegraphics[width=0.85\textwidth]{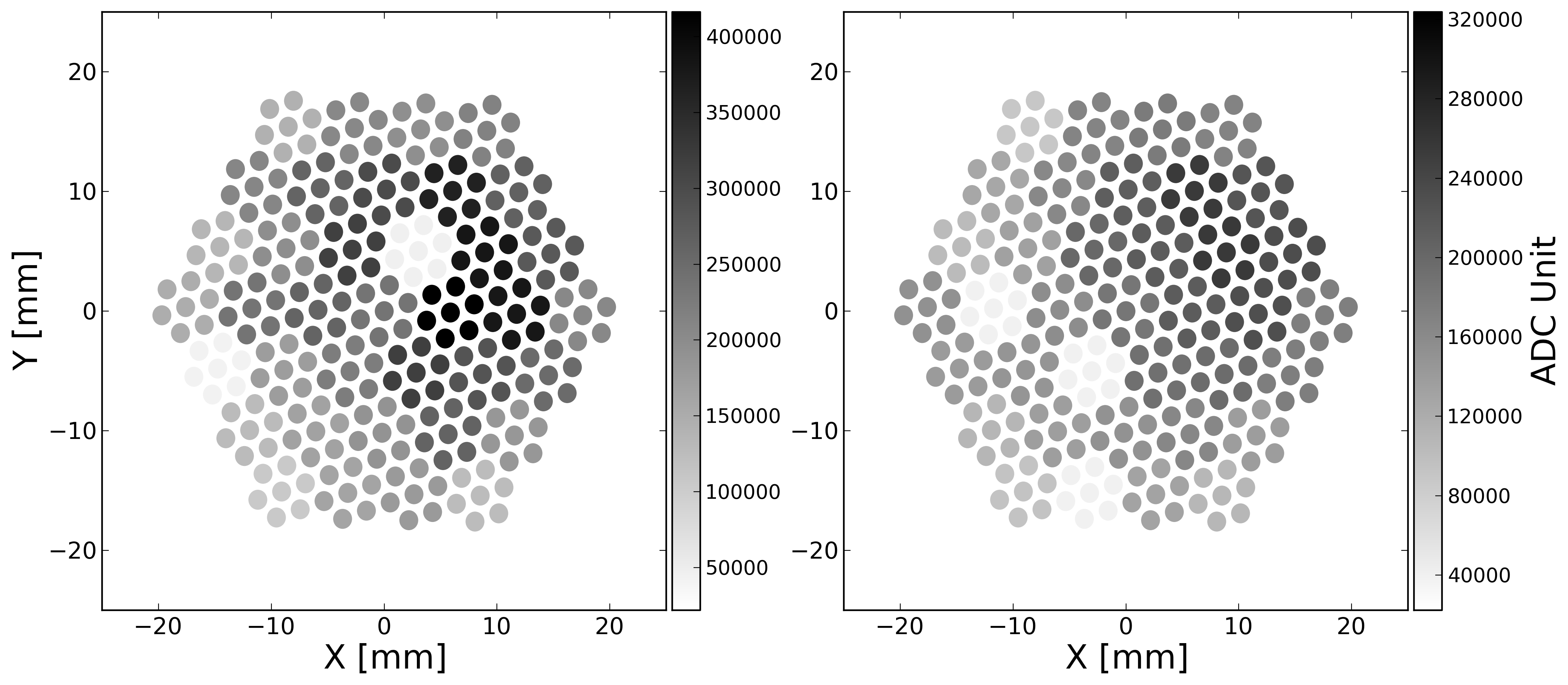}
    \caption{APD channels' luminosity of GAN-generated waveforms.}
    \label{fig:dead_channel_sub1_B}
    \end{subfigure}\hspace{1em}%
    \begin{subfigure}{.95\textwidth}
    \centering
    \includegraphics[width=0.85\textwidth]{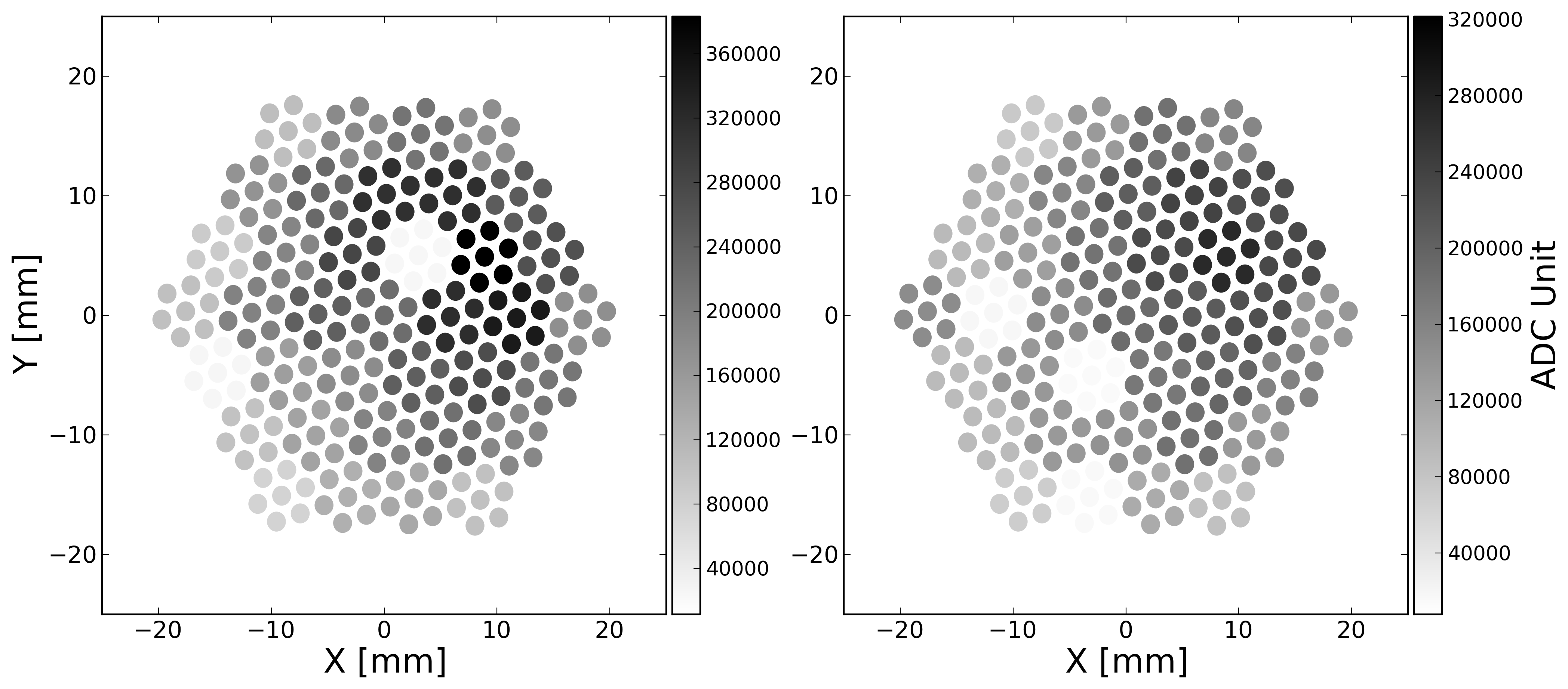}
    \caption{APD channels' luminosity of real waveforms.}
    \label{fig:dead_channel_sub2_B}
    \end{subfigure}\hspace{1em}%
    \caption{Summed amplitude of all events in a run for each of the APD channels. GAN-generated (top) and real (bottom) data are shown. Two APD planes are shown separately. Darker color means higher amplitude. White areas show the dead channels. We can identify the disconnected channels in GAN data (a) that matches with calibration data in (b).} 
    \label{fig:dead_channel}
\end{figure}

Another feature that is easy to spot is that several waveforms do not have signals. These correspond to five channels disconnected due to excessive leakage current of the APDs. We can examine the disconnected channels further by looking at the average signal amplitude of all events in one typical calibration run. Fig.~\ref{fig:dead_channel} shows the pattern of APDs in two planes and its corresponding signal amplitude for both GAN and real waveforms. The APDs are arranged in two hexagonal planes and grouped in gangs of 6 or 7. The amplitudes are represented in grey-scale in the plot with darker color representing larger amplitudes.  
As shown in Fig.~\ref{fig:dead_channel_sub1_B} and \ref{fig:dead_channel_sub2_B}, the model finds the disconnected APD channels and reproduces an average amplitude pattern similar to the real calibration data. The average amplitude of the disconnected APDs in the generated waveforms does not completely go to zero, as some GAN-generated events have small residual pulses in these channels. The average amplitudes are around 5 ADC units slightly larger than the noise RMS value of 2 ADC units, but negligible compared to amplitudes of connected channels.  

\begin{figure}
    \centering
    \begin{subfigure}[t]{.4\textwidth}
        \centering
        \includegraphics[width=\linewidth]{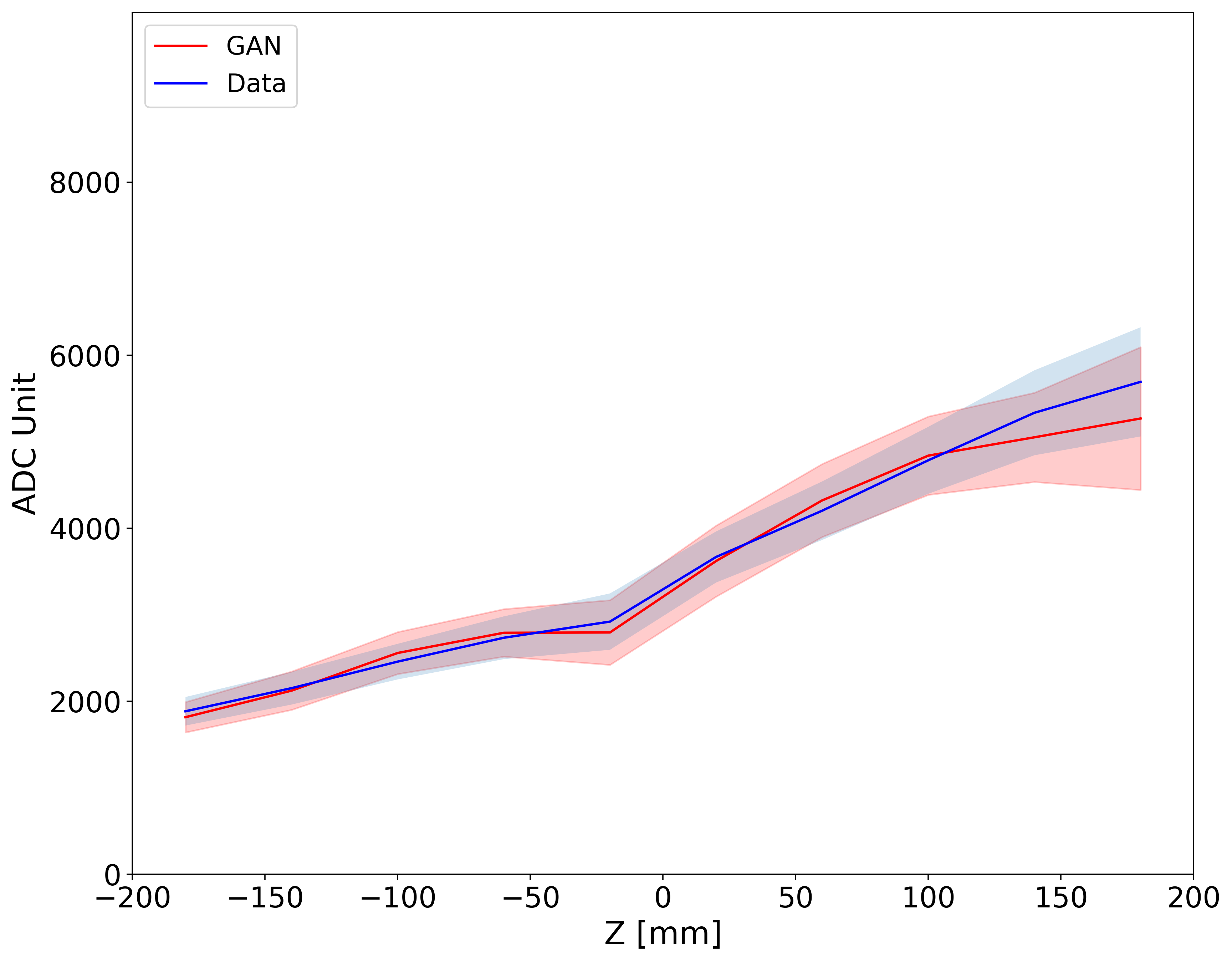}
        \caption{Plane 1 Amplitude vs. Z}
        \label{fig:zmap_1_B}
    \end{subfigure}\hspace{1em}%
    \begin{subfigure}[t]{.4\textwidth}
        \centering
        \includegraphics[width=\linewidth]{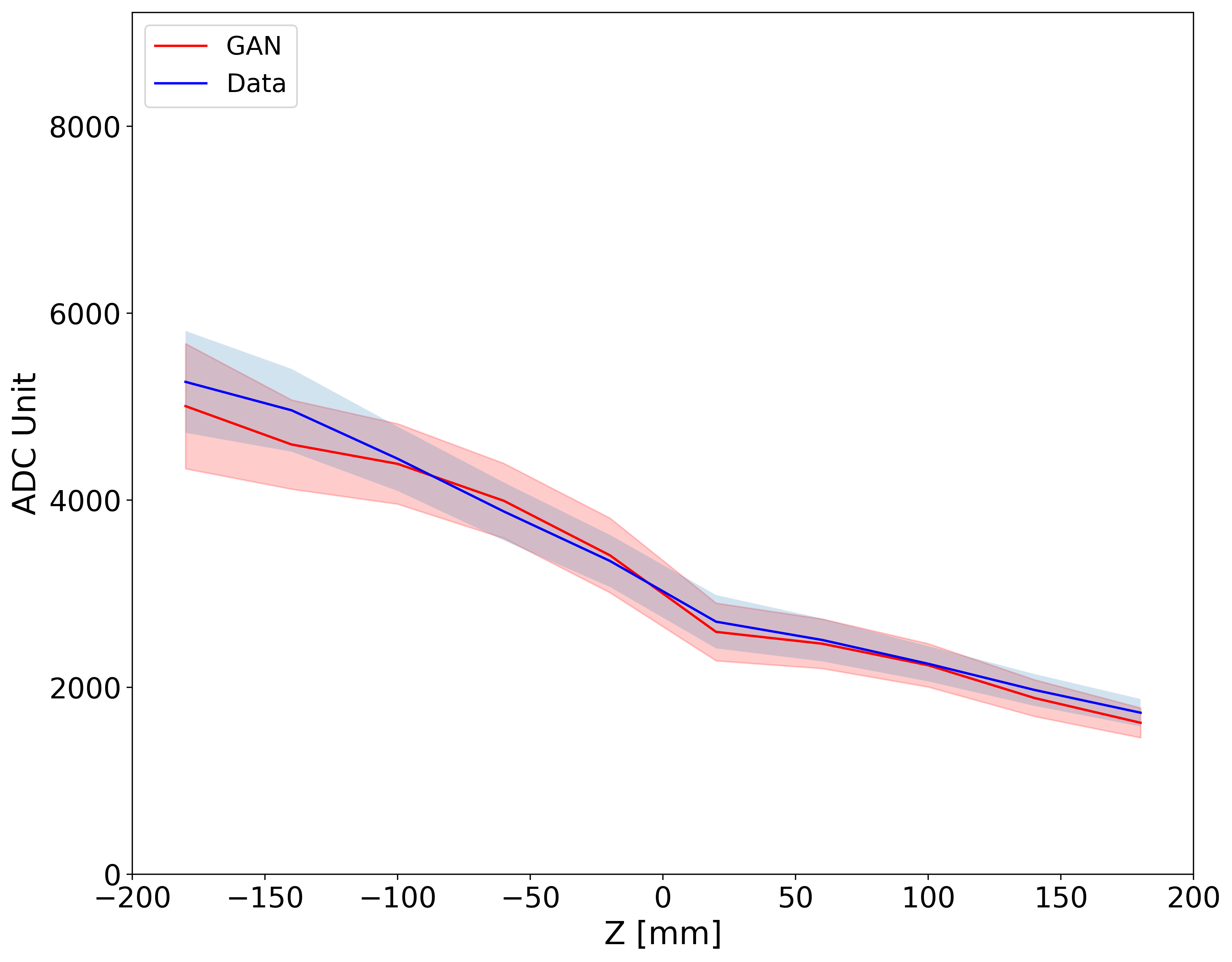}
        \caption{Plane 2 Amplitude vs. Z}
        \label{fig:zmap_2_B}
    \end{subfigure}\hspace{1em}%
    \begin{subfigure}[c]{.4\textwidth}
        \centering
        \includegraphics[width=\linewidth]{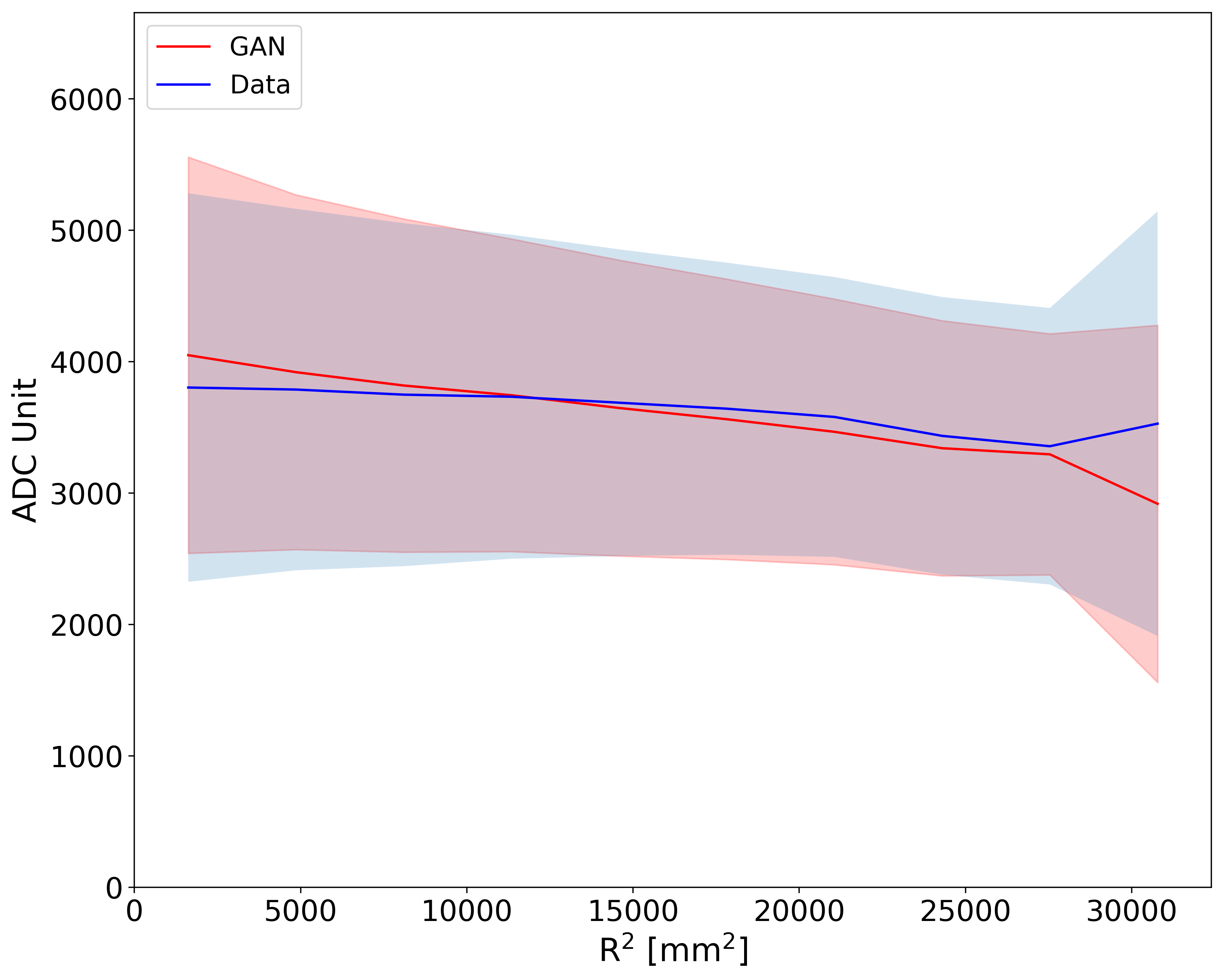}
        \caption{Plane 1 Amplitude vs. $\mathrm{R^2}$}
        \label{fig:rmap_1_B}
    \end{subfigure}\hspace{1em}%
    \begin{subfigure}[c]{.4\textwidth}
        \centering
        \includegraphics[width=\linewidth]{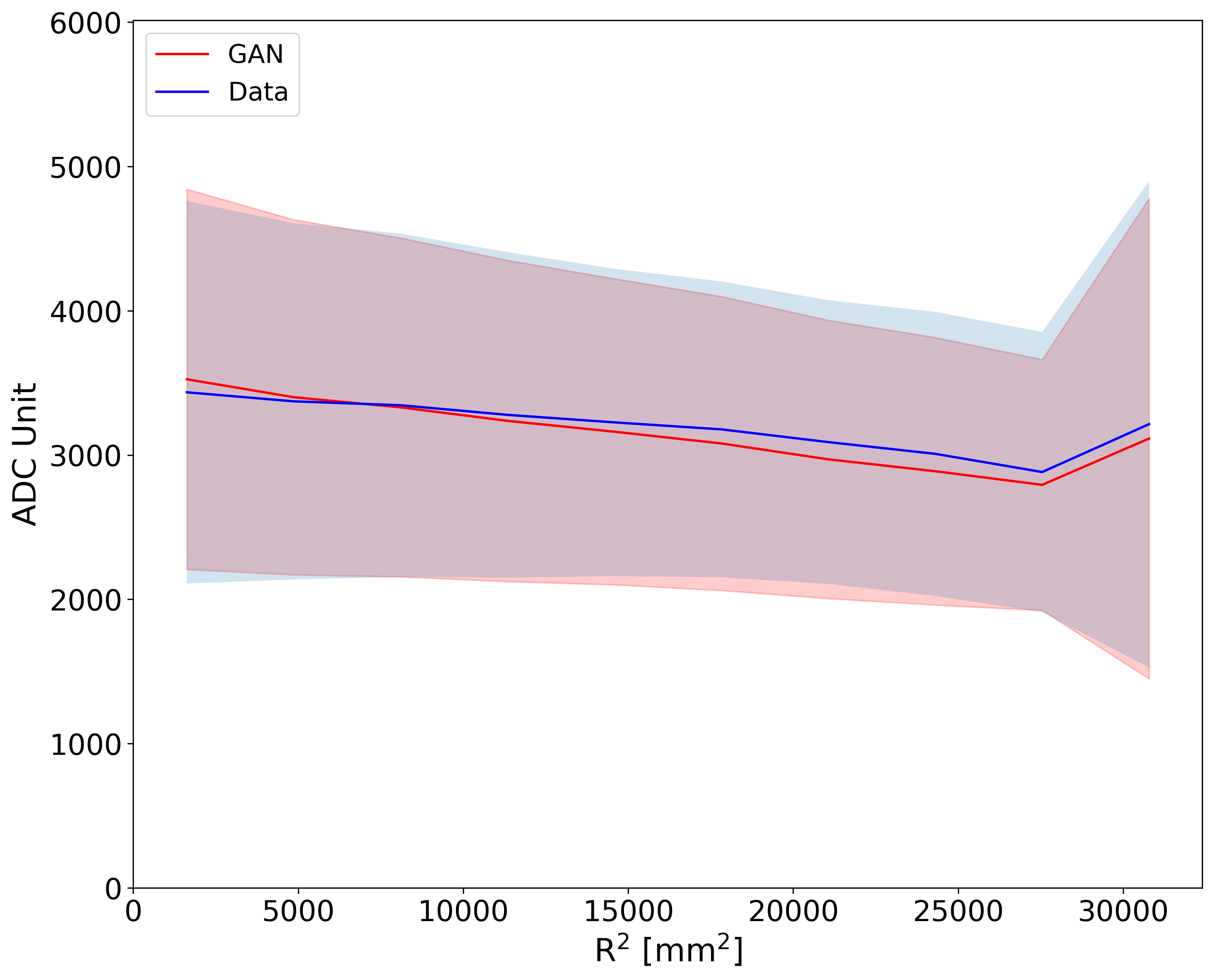}
        \caption{Plane 2 Amplitude vs. $\mathrm{R^2}$}
        \label{fig:rmap_2_B}
    \end{subfigure}\hspace{1em}%
    \begin{subfigure}[b]{.4\textwidth}
        \centering
        \includegraphics[width=\linewidth]{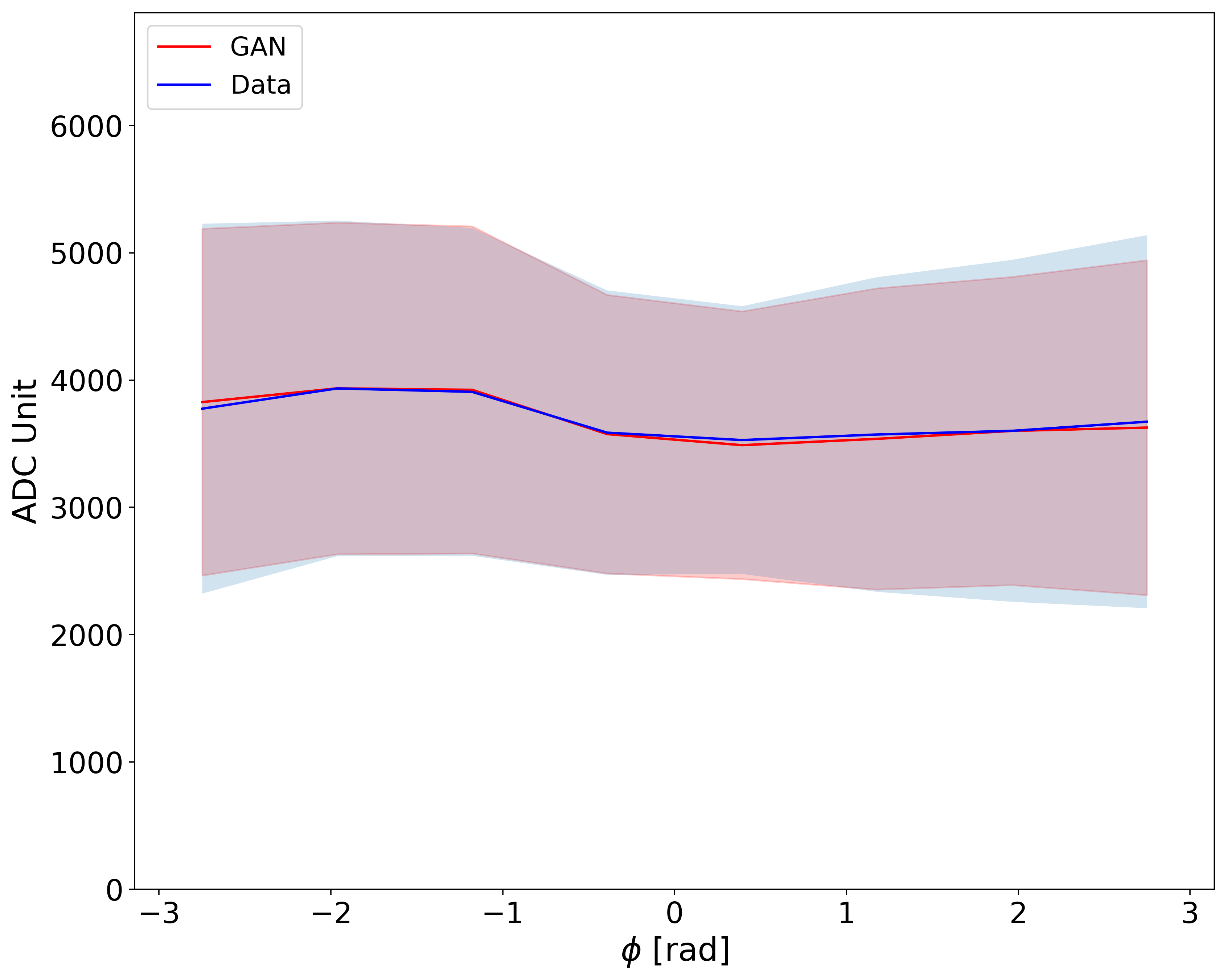}
        \caption{Plane 1 Amplitude vs. $\mathrm{\phi}$}
        \label{fig:phimap_1_B}
    \end{subfigure}\hspace{1em}%
    \begin{subfigure}[b]{.4\textwidth}
        \centering
        \includegraphics[width=\linewidth]{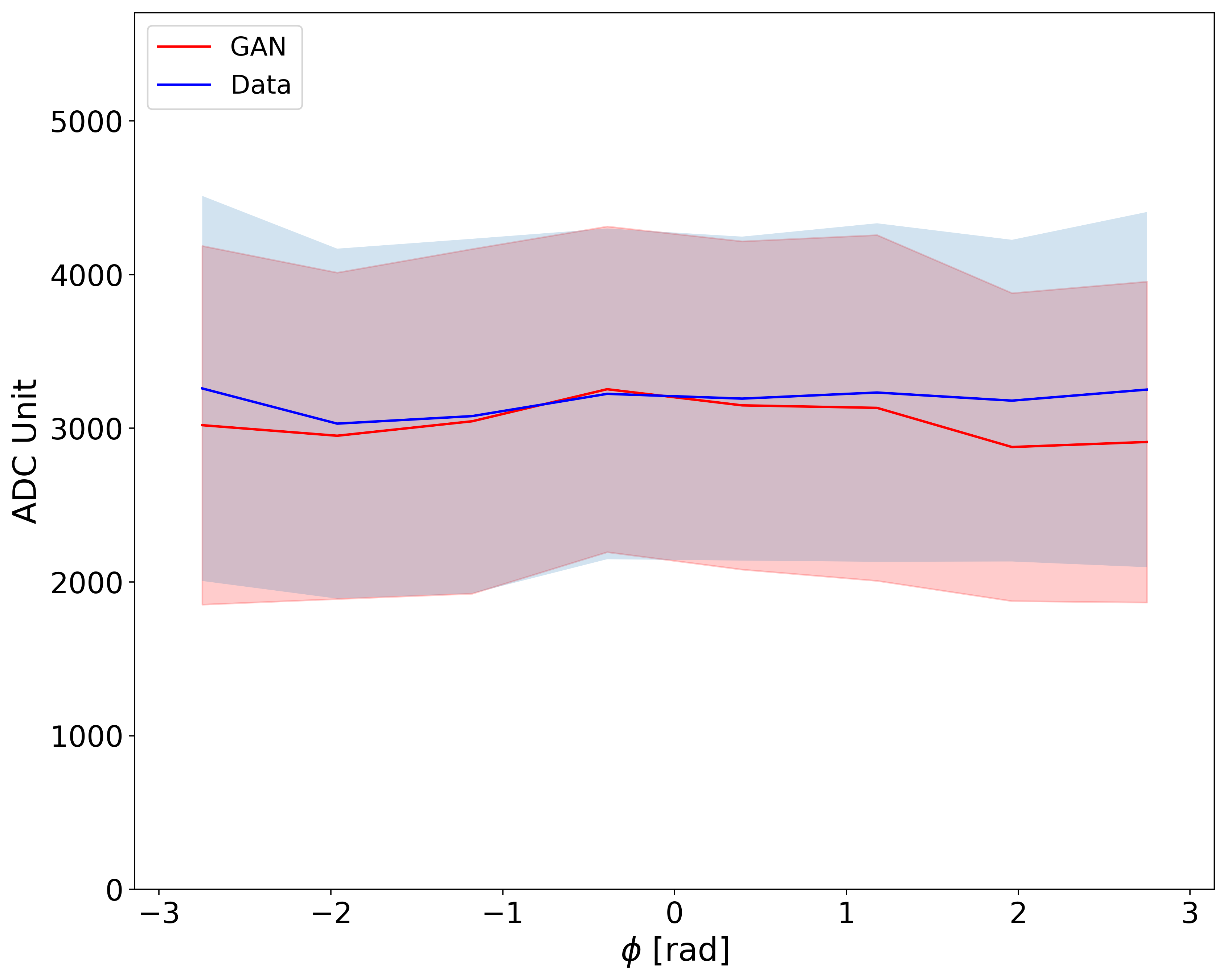}
        \caption{Plane 2 Amplitude vs. $\mathrm{\phi}$}
        \label{fig:phimap_2_B}
    \end{subfigure}\hspace{1em}%
    \caption{Profiles of the validation metric distributions as the function of the spatial components. For each metric $i$, we show its average value $\mu_i$ (middle thick line), as well as the one standard deviation band of the metric distribution $\mu_i \pm \sigma_i$ (top and bottom band edge). GAN-generated distributions (red) show a strong overlap with real data (blue).}
    \label{fig:amplitude_map}
\end{figure}

To make a more precise evaluation, we apply the EXO-200 reconstruction algorithm to both GAN-generated waveforms and the real data. Since the reconstruction requires wire signals to determine the position of the events, we copied the wire signals of the real data to the generated APD waveforms. Since the wire signals are identical, the reconstruction returns the same position and charge energy information for both the GAN and real data. The only difference comes from the reconstructed scintillation energy. To extract the scintillation energy, the reconstruction algorithm adds all waveforms from plane 1 and plane 2 to produce two sum waveforms, then fits the sum waveforms to extract the summed amplitudes. In Fig. \ref{fig:amplitude_map}, we plot the spatial dependence of the sum waveform amplitude of each APD plane for both real data and data generated with the energy and position as labels. Plane 1 is located at z = 200 mm while plane 2 is located at z = -200 mm, thus the plane 1 amplitude becomes higher when a cluster is closer to the plane 1 while the plane 2 amplitude becomes lower. There is no large $R$ or $\phi$ dependence due to high reflectivity of the PTFE reflector surrounding the TPC. The plot shows that GAN accurately reproduces the plane 1 and plane 2 average amplitudes and that standard deviation matches well with the real data in all three cylindrical spatial dimensions. 

After signal fitting, the extracted amplitudes go through energy calibration procedure to convert them into scintillation energy measurements.  
\begin{figure}
    \centering
    \begin{subfigure}[t]{0.45\textwidth}
        \centering
        \includegraphics[width=\linewidth]{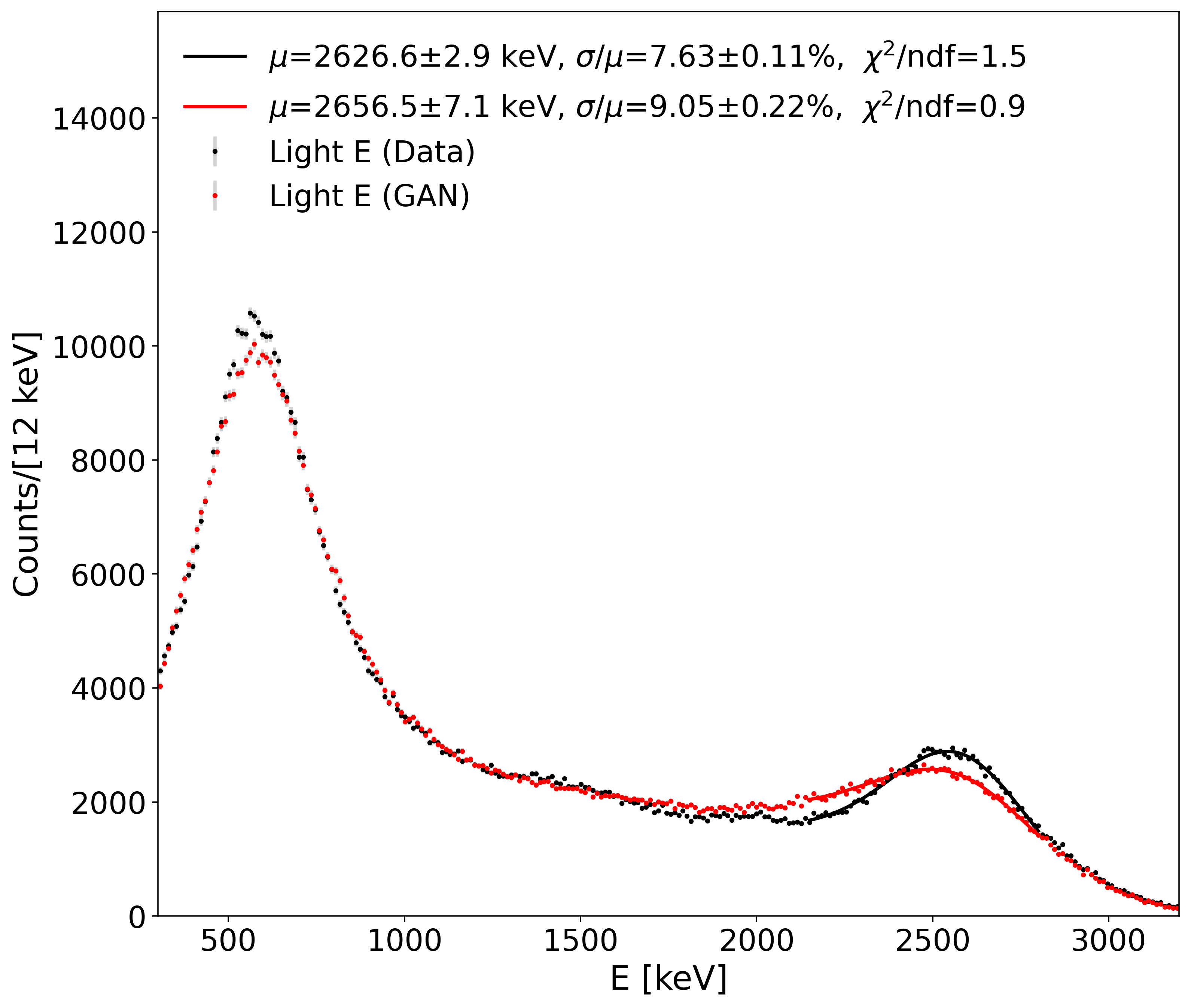}
        \caption{The $^{228}$Th light energy spectra of real (black) and GAN-generated (red) events without position correction. Fitted resolutions are also given.}
    \label{fig:NoPosCorr_rosolution}
    \end{subfigure}\hspace{1em}
    \begin{subfigure}[t]{0.45\textwidth}
        \centering
        \includegraphics[width=\linewidth]{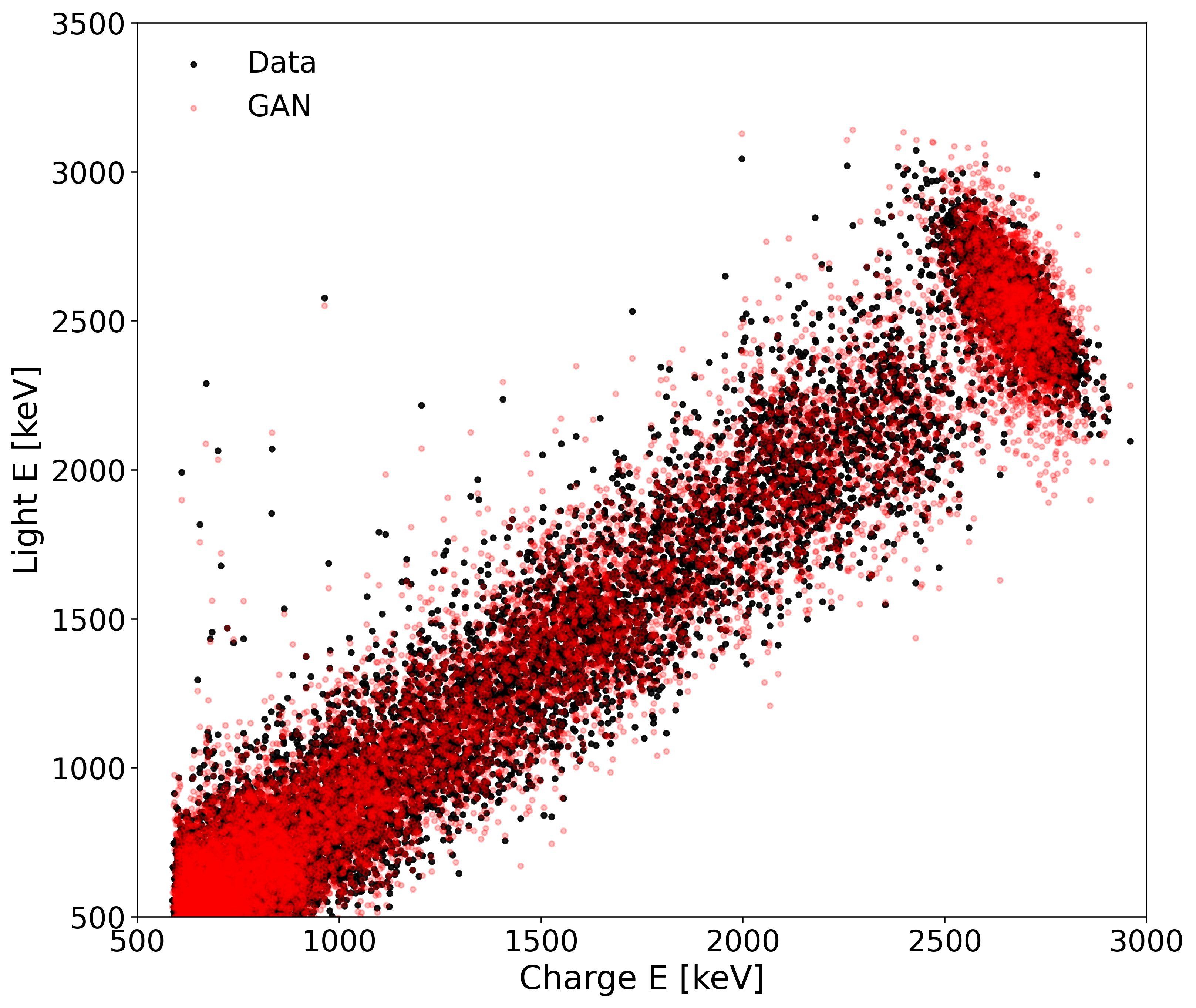}
        \caption{Scatter plot of reconstructed light and charge energies from GAN (red) and real (black) waveforms. Both GAN and real data show anti-correlation between the light and charge signals.}
        \label{fig:anti-correlation}
    \end{subfigure}\hspace{1em}
    \begin{subfigure}[b]{0.45\textwidth}
        \centering
        \includegraphics[width=\linewidth]{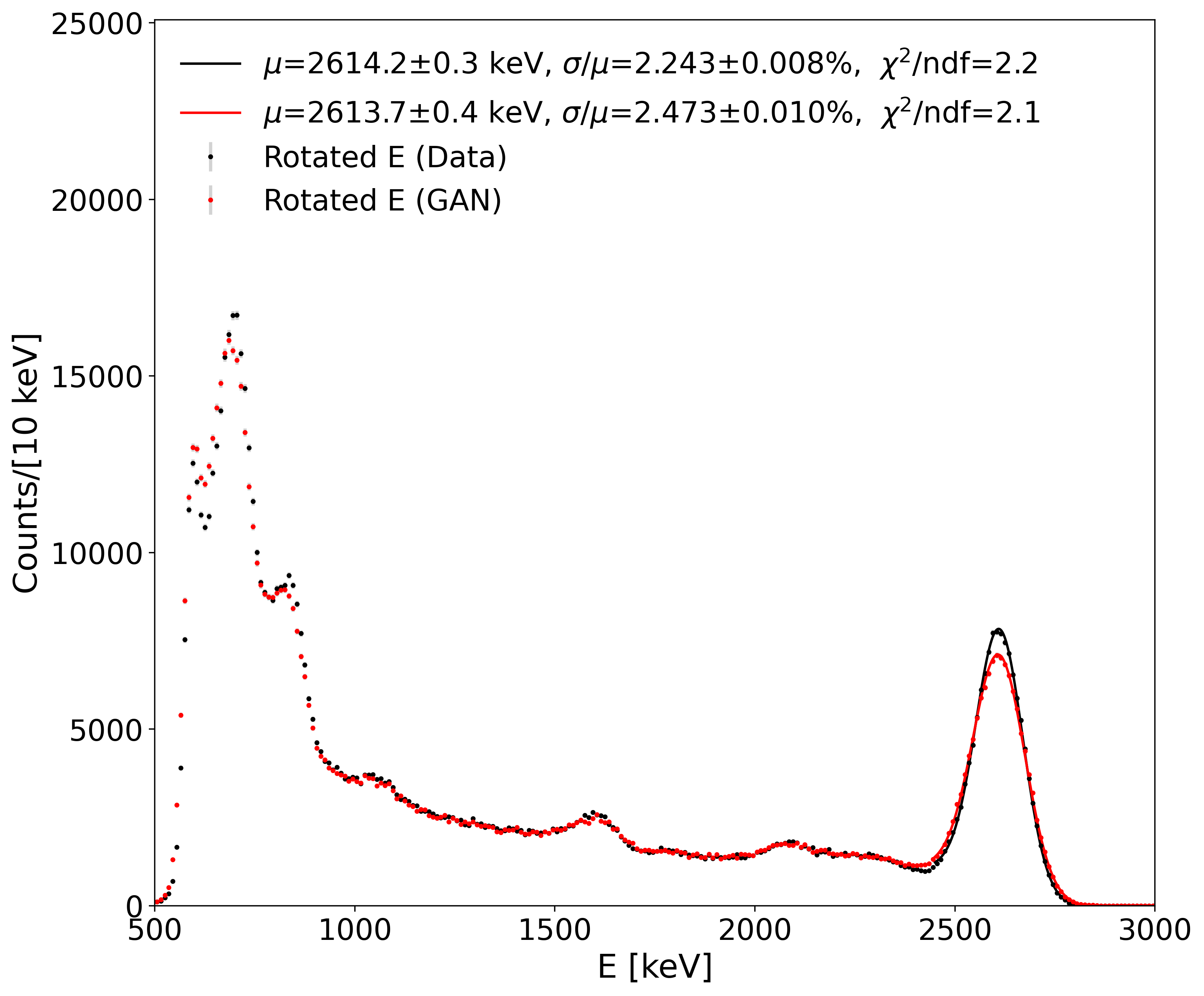}
        \caption{Histograms show the projection of the scatter points from Fig. \ref{fig:anti-correlation} onto the corresponding rotated axis, $\theta_\mathrm{r}$ = 0.102 and $\theta_\mathrm{g}$ = 0.082. The red and blue lines show the fitting results of the rotated E peak.}
        \label{fig:anti-correlation-projection}
    \end{subfigure}\hspace{1em}
    \begin{subfigure}[b]{0.45\textwidth}
        \centering
        \includegraphics[width=\linewidth]{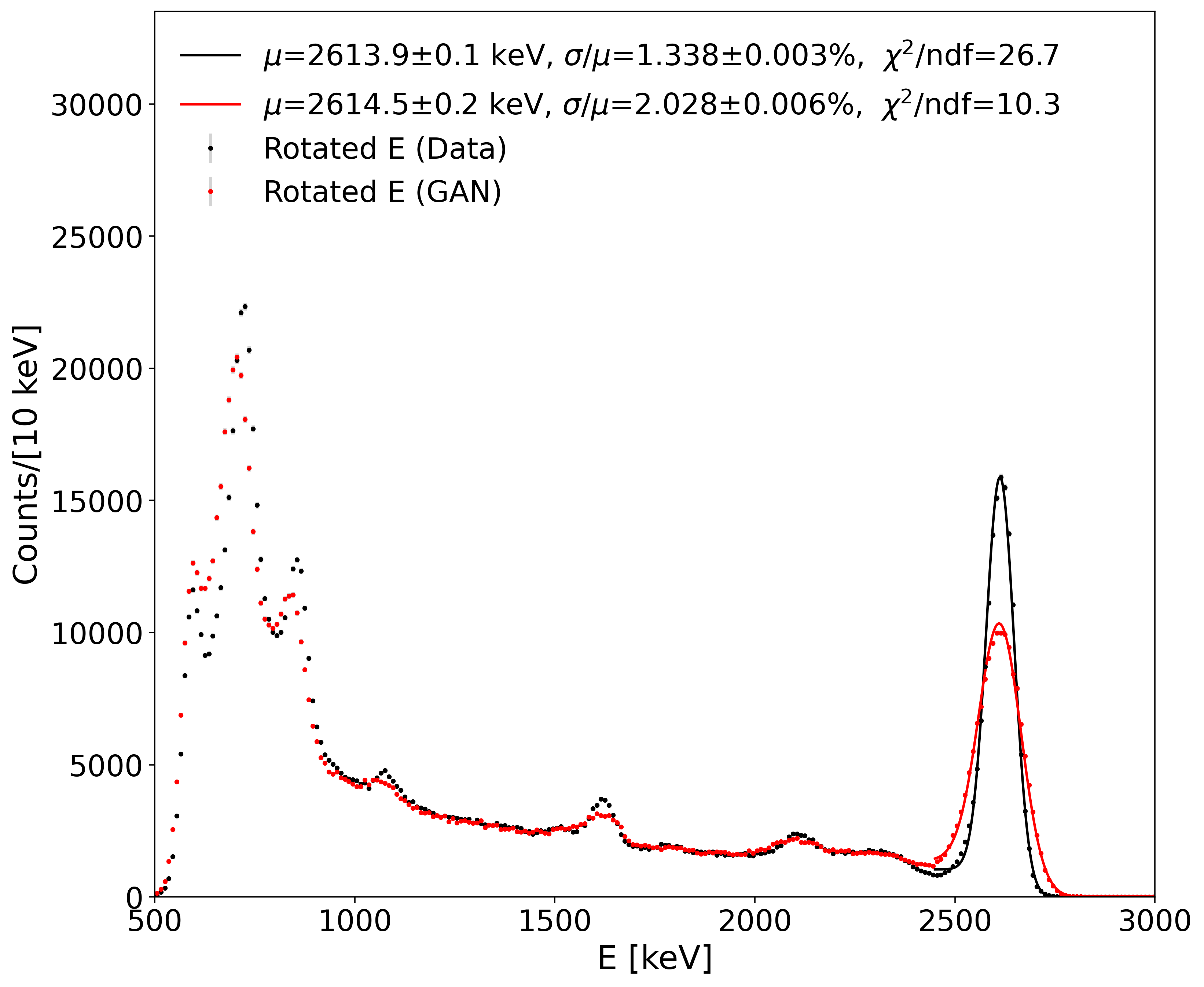}
        \caption{Histograms show the same projection of events after the light map calibration onto corresponding rotated axis, $\theta_\mathrm{r}$ = 0.224 and $\theta_\mathrm{g}$ = 0.143. The red and blue lines show the fitting results of the rotated E peak.}
        \label{fig:anti-correlation-projection-lightmap2017}
    \end{subfigure}\hspace{1em}
    \caption{Reconstructed light energy distributions for the GAN-generated and real data.}
\end{figure}
The scintillation spectra of the generated and real data for one $^{228}$Th calibration run are shown in Fig.~\ref{fig:NoPosCorr_rosolution}. The agreement is fairly good across the entire energy range, but the GAN generated spectrum has slightly worse energy resolution. The energy resolution ($\sigma$/E) of the 2.614 MeV $^{228}$Th peak is 7.6\% for real data, and 9.1\% for GAN-generated data. The degradation of the resolution is likely a consequence of the uncertainty of the energy labels used to train the network, rather than the inherent issue with the GAN network. 

As mentioned in Section~\ref{sec:tpc}, using a linear combination of the charge and light signals can improve the energy resolution of the detector. In a 2-D scatter plot of scintillation energy and charge energy, this corresponds to a projection to a rotated axis, as shown in Fig.~\ref{fig:anti-correlation}. The GAN-generated data reproduces the anti-correlation between the charge and light signals, confirming that the good agreement between true and generated scintillation energies is achieved on the event-by-event basis. The optimal rotation angle, which in EXO-200 was treated as a free parameter chosen to optimize rotated resolution, is slightly different from the real data. If we project the GAN and real datasets onto their optimal axes, the rotated spectra match well with each other, as shown in Fig.~\ref{fig:anti-correlation-projection}. The energy resolution of the $^{228}$Th peak is 2.24\% for the real data and 2.47\% for the GAN generated data. 

The energy resolution can be further improved by correcting position dependency of the light response. To that end, the detector is divided into cylindrical voxels with 13 sections in $Z$ and $R$ directions and 8 sections in $\phi$ direction. The $^{228}$Th calibration data in each voxel is fitted to extract the position of the 2.614 MeV peak. The ratio between the voxel peak position and the overall peak position is the correction factor for the voxel. We call this the light map correction~\cite{improved_2nu}. The light map records the correction factor in the reconstruction space. If we apply the light map to both the real and GAN generated data, the energy resolution improves to $\sim$1.3\% for the real data, and $\sim$2.0\% for the GAN data, as shown in Fig. \ref{fig:anti-correlation-projection-lightmap2017}. The fact that the light map correction improves the GAN resolution demonstrates that GAN network is able to reproduce, at least to a good degree, the position dependence of the scintillation signal. Like in the previous two tests, the worse resolution for the GAN images is likely also impacted by the imperfect energy labels used in training. If so, then using a light source with known intensity adjustable to span the relevant energy range and deployed at relevant positions inside the detector~\cite{sno_cal,db_cal,dc_cal,borexino_cal,juno_cal} to train the network would lead to better results.  

To compare the simulation speed, we generate scintillation waveforms for several thousands $^{228}$Th events inside the detector using the EXO-200 Monte Carlo framework and GAN network. 
The EXO-200 framework first utilizes Geant4 to simulate energy depositions of the $^{228}$Th gammas, then uses a fast parametric optical simulation that produces expected number of photons detected by the APD planes, and finally generates the APD waveforms using transfer functions that depend on the parameters of the APD electronics. By default, an analogous ``digitization'' process is performed for the wires, but we switch the wire digitization off for a more fair comparison. We observe an average simulation rate of $\sim$4.2 events/s, with the Geant4 and digitization steps taking roughly the same time. The test was run on a machine with an Intel Xeon Gold 6226R CPU~\cite{xeon_gold} and 12 GB RAM. In contrast, GAN directly simulates APD waveforms given $^{228}$Th event energy and position. We observe an average simulation rate of $\sim$69 events/s when ran on Nvidia GeForce 1080Ti GPU~\cite{geforce} with 11 GB RAM. While not a precise back-to-back comparison, the test shows that the new approach can simulate light waveforms at a rate that is roughly an order of magnitude faster than with the traditional approach. 

%%%%%%%%%%%%%%%%%%%%%%%%%%%%%%%%%%%%%%%%%%%%%%%%%%%%%%%%%%%%%%%%%%

\section{Conclusions}
%\label{sec:conclusion}
In this work, we applied Wasserstein GAN technique to generate raw APD signal waveforms for the EXO-200 experiment. Detector calibration data was used as the training samples bypassing the need for computationally intensive scintillation simulation and inaccuracies of the MC simulation models. Using reconstructed event position and energy as labels, the network was successfully trained to generate raw waveforms that mimic main features of the real waveforms, including signal arriving time, pulse shape, and channel noise. At the signal level, the network learned the missing channels and spatial dependency of the pulse amplitudes. Furthermore, the GAN could reproduce the energy spectrum of the calibration data after reconstruction. Combined with corresponding charge waveforms, the anti-correlation between the scintillation and charge channels was reproduced, confirming that the good agreement between true and generated scintillation energies is done event-by-event. The energy resolution of the scintillation and rotated spectra are slightly worse for the GAN-generated data than for the real data. 
%%%%%% Disscussion about the multi-cluster events %%%%%%%%%%%

In this study, events that contain single spatially-distinct charge deposition (so-called single-site events) for both training and validation were used. Although there are also multi-site events in the dataset, the only difference between them as far as this work is concerned is the spatial distribution of the scintillation light. Since the GAN was able to reproduce spatial distribution of the signal for single-site events, it is reasonable to expect a similar performance on multi-site events, which are effectively a combination of several independent sites. We concentrated on the single-site events for simplicity in this novel work, leaving the multi-site events as a potential future direction. It should also be noted that since the approach described here requires training on real data, it can not be used to optimize the design of future detectors as effectively as the traditional MC frameworks. 
%%%%%%%%%%%%%%%%%%%%%%%%%%%%%%%%%%%%%%%%%%%%%%%%%%%%%%%%%%%%%%%

We have demonstrated that using the GAN network and calibration data can be a powerful approach to generate simulation data faster and more accurately than with traditional Monte Carlo simulation. One drawback of the approach is that the labels of the training data are derived from reconstructed scintillation energy, which in case of EXO-200 carries a $\sim$5\% uncertainty. Experiments that can utilize a set of training events with better truth labels would see better results. For example, some experiments may be able to use an LED or laser source with known, variable intensity, and which can illuminate all relevant parts of the detector. An interesting alternative possibility, which we also consider an avenue for future work, is training the network on both charge and light waveforms simultaneously. If the network is able to deduce the anti-correlation between the two signals, then one could potentially harness the advantage of the improved accuracy of the rotated energy labels.

In summary, GAN is a promising tool to accelerate and improve simulation for particle physics. Using calibration data directly as inputs can simplify the simulation and avoid inaccuracies of the simulation model.

\acknowledgments
This work is supported by a Department of Energy Grant No. DE-SC0019261. EXO-200 is supported by DOE and NSF in the United States, NSERC in Canada, IBS in Korea,  DFG in Germany, and CAS and ISTCP in China. EXO-200 data analysis and simulation uses resources of the National Energy Research Scientific Computing Center (NERSC). We gratefully acknowledge the support of Nvidia Corporation with the donation of three Titan Xp GPUs used for the optical simulations.

\bibliographystyle{JHEP} 
\bibliography{main}

\appendix
\section{Network Architecture}
\label{appendix_A}
\begin{figure}[htpb]
    \centering
    \includegraphics[width=\linewidth]{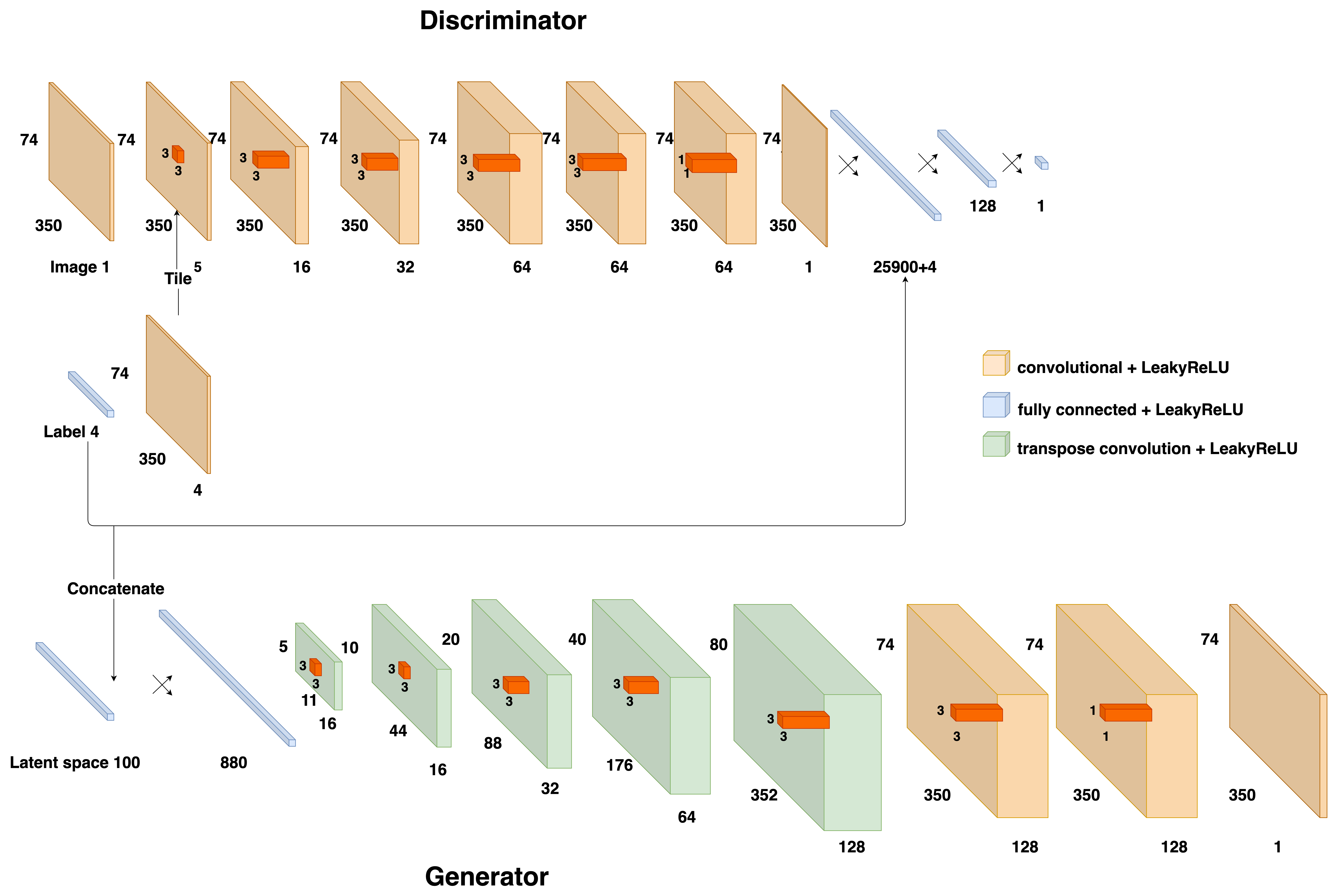}
    \caption{Architecture of the GAN with corresponding kernel size, number of feature maps. The discriminator (generator) has 1,755,333 (1,273,953) trainable parameters.}
    \label{fig:GAN_networks}
\end{figure}

The neural network architectures for the discriminator and generator are shown in \ref{fig:GAN_networks}.  
\end{document}